\documentclass[journal]{IEEEtran}
\usepackage{cite}
\usepackage{amsmath,amssymb,amsfonts}
\usepackage{algorithm}
\usepackage{algorithmic}
\usepackage{graphicx}
\usepackage{booktabs,tabularx,array}
\usepackage[mathlines,switch]{lineno}
\usepackage{multicol,multirow}
\usepackage{marginnote}
\usepackage{stfloats}
\usepackage{colortbl,hhline}

\title{NAS Timers for Non-Terrestrial Networks:\\Current Standards and Directions} %Note: This is a draft title
\author{{Cheng Liu, and Peng Hu, \textit{Senior Member,~IEEE}}
\thanks{Cheng Liu and Peng Hu are with the Department of Electrical and Computer Engineering, University of Manitoba, Winnipeg, MB, Canada R3T 2N2 (e-mail: liuc13@myumanitoba.ca, peng.hu@umanitoba.ca).}
}

\begin{document}

\maketitle

\begin{abstract}
Response-supervision timers in the 5G Non-Access Stratum (NAS) bound the time allowed for signaling exchanges before retransmission or abnormal-case handling begins. In non-terrestrial networks, that interval can include satellite propagation, terrestrial transport, and processing by multiple core-network functions. Under current 3GPP specifications, the applicable timer value for satellite access is selected according to the satellite NG-RAN radio access technology (RAT) type. However, control paths within the same low Earth orbit (LEO) network can differ in propagation delay and processing load. This article reviews the relevant 3GPP timer semantics and traces the delay components covered by different timers through 5G registration. It organizes candidate refinements along two dimensions: configuration granularity and information scope. A registration case study illustrates the trade-off between avoiding premature timer expiry and recovering promptly from message loss. The article identifies the information and coordination that future specifications would need to address to support interoperable timer configuration.
\end{abstract}

\section{Introduction}\label{sec:I}
Non-terrestrial networks (NTNs) have progressed from feasibility studies to integration within the 5G system (5GS), with low Earth orbit (LEO) constellations offering lower propagation delay than higher-orbit satellite systems \cite{Lin2021_5GFromSpace}. This integration places established control-plane procedures on communication paths whose timing characteristics depend on both orbital geometry and network architecture. Non-Access Stratum (NAS) signaling supports registration, service request, deregistration, authentication, and security mode control between the user equipment (UE) and the core network \cite{3gpp_ts23502}. Across these procedures, timers determine when an incomplete message exchange is retransmitted, terminated, or handled according to the applicable abnormal-case procedure.

NAS timers encode several kinds of protocol requirements. Response-supervision timers bound the interval during which an endpoint waits for an expected message. Backoff timers regulate subsequent access attempts, whereas periodic-update and validity timers maintain policy or protocol state over longer intervals. Path delay most directly affects response-supervision timers because their expiry is evaluated against the duration of a protected exchange. Timer function, ownership, and the protected exchange therefore determine which NTN delay components are relevant to timer configuration.

\begin{figure*}[!t]
\centering
\includegraphics[width=0.92\textwidth]{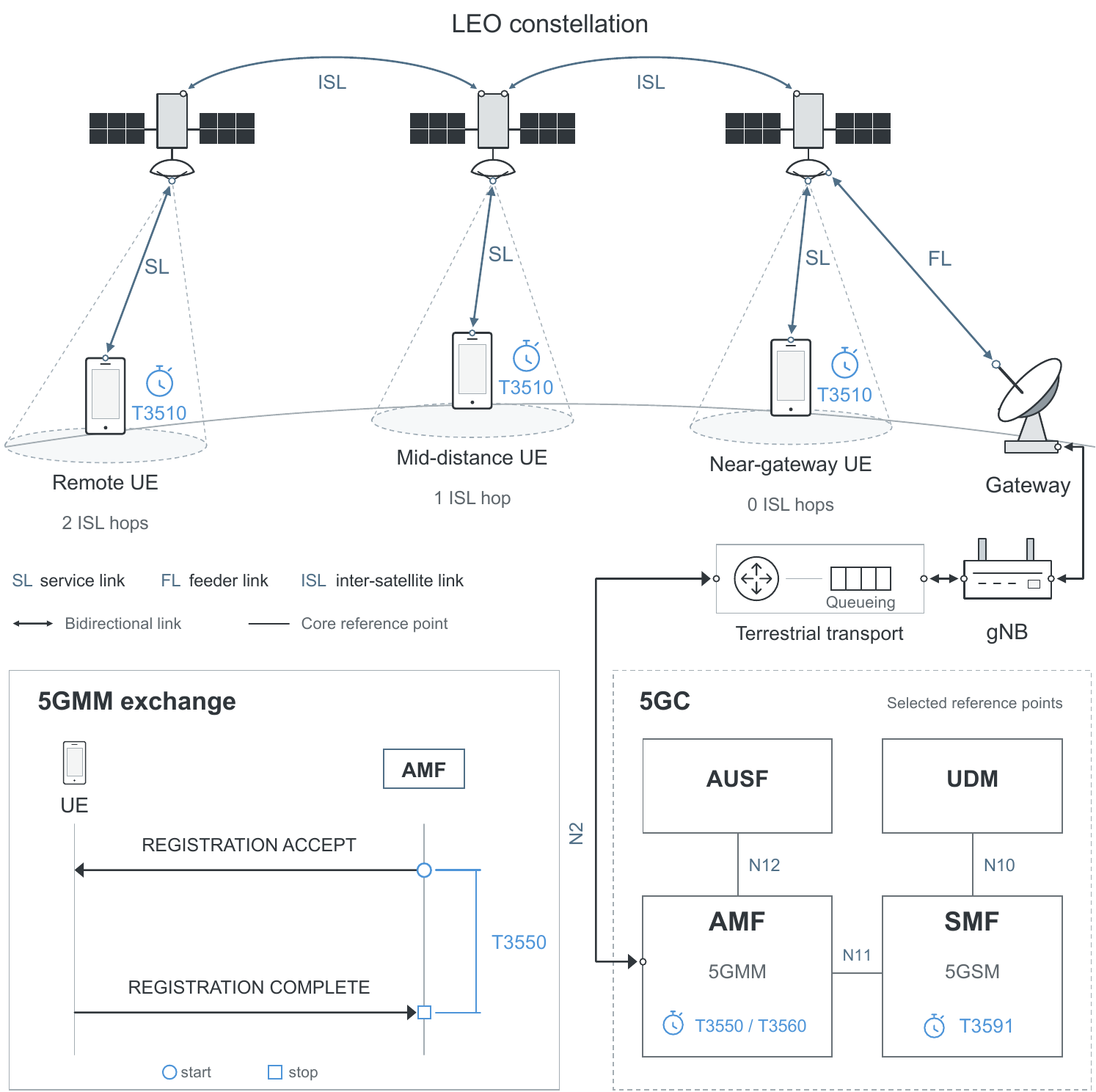}
\par\vspace{7pt}
\caption{Illustrative LEO-integrated 5GS control path and NAS timer placement.}
\label{fig:ntn-nas-overview}
\end{figure*}

The end-to-end delay observed by a NAS timer can differ substantially from the propagation delay of the radio link. A protected exchange may traverse a service link, one or more inter-satellite links (ISLs), a feeder link, a gateway, terrestrial transport, and several core-network functions \cite{3gpp_ts38300}. Fig.~\ref{fig:ntn-nas-overview} shows a reference deployment with a ground-based gNB and 5G Core (5GC) functions, where satellite relay paths traverse different numbers of ISLs before reaching a shared gateway. These paths show how UEs within one LEO constellation can encounter different delays under the timer value associated with the same satellite radio access technology (RAT) type. Satellite motion, route selection, gateway attachment, queueing, and network-function load can change the accumulated end-to-end delay, whereas timer ownership and the protected exchange determine which components fall within the timer interval. A short timer may expire while a valid response is still in transit or awaiting processing, causing avoidable retransmissions or premature procedure failure. An excessively long timer delays loss detection and retains procedure state after successful completion has become unlikely.

Current 3GPP specifications provide an interoperable baseline through predefined timer values and timer selection based on the satellite NG-RAN RAT type for satellite access. The applicable timer value is selected when a procedure starts and remains unchanged for that procedure instance \cite{3gpp_ts24501}. Keeping the selected value unchanged preserves stable timer semantics even if route or mobility conditions change during the procedure. The satellite NG-RAN RAT type identifies the applicable satellite access category. The complete control path also reflects the ISL route, gateway and core-network placement, queue exposure, and processing functions traversed by a particular NAS exchange.

This distinction exposes several open issues for NTN timer design. UEs within the same satellite RAT type can experience different control-path delays, while UE-owned and network-owned timers observe different subsets of propagation and processing time. Information relevant to timer selection may also reside at an entity other than the timer owner. Mobility can change the applicable access or route conditions during an ongoing procedure, yet a running timer must retain predictable semantics. Any refinement must also coexist with legacy endpoints and preserve the standardized start event, stop event, expiry action, and fallback behavior. The standards problem is thus to associate available context with the appropriate timer owner without compromising interoperability.

Research relevant to this problem spans several protocol layers. Adaptive timer and buffer configurations at the Radio Link Control and Packet Data Convergence Protocol layers address long or location-dependent round-trip times \cite{Sheemar2023_AdaptiveTimers,Sheemar2024_LocationTimers}. Work on NTN mobility \cite{Liu2026_NTNMobility} addresses path continuity, while Access and Mobility Management Function (AMF) scalability mechanisms address core-network load \cite{Alawe2018_AMFScalability}. At the NAS layer, AstroTimer \cite{AstroTimer2026} illustrates how satellite propagation and network-function processing can inform timer configuration. These strands motivate a broader design space comprising condition-based selection among predefined values, endpoint-local calculation, network-assisted selection, mobility-consistent handling, capability negotiation, and fallback. Path-aware configuration occupies one part of this design space and relies on information that reaches the timer owner before selection.

This article relates the current 3GPP timer framework to delay along heterogeneous NTN control paths. It maps representative registration timers to their owners and protected exchanges, identifies where propagation and processing enter each interval, and organizes fine-grained NAS timer configuration by configuration granularity and information scope. A registration case study compares timer behavior across fixed and finer-grained configurations. The article concludes with implications for interoperability, assistance mechanisms, and future specification work.

\section{NAS Timer Standardization in 3GPP}

3GPP support for NAS timer operation over satellite access has progressed from NTN studies to normative satellite-specific timer values and RAT-type-based selection rules. The resulting framework defines the timer ownership, procedure boundaries, and values that govern each protected exchange.

%<AUG 27, 2026>NOTE: a table can be used if you'd like to compare with different 3GPP releases for their features regarding the NAS timer. 

\subsection{Evolution and Scope of 3GPP NTN Standardization}

3GPP NTN standardization progressed from Rel-15 and Rel-16 studies to normative NR NTN support in Rel-17 \cite{3gpp_tr38811,3gpp_tr38821}. In Rel-17, TS 24.501 introduced satellite-specific NAS timer values and selection based on the satellite NG-RAN RAT type, followed by targeted corrections and clarifications in Rel-18 and Rel-19. According to 3GPP TS 38.413 from Rel-17 \cite{3gpp_ts38413}, a UE may access a 5G network through satellite-based NR access. The satellite access type is identified by the satellite's orbit as NR(LEO), NR(MEO), or NR(GEO), corresponding to low-, medium-, and geostationary-Earth-orbit satellites, respectively. Table~\ref{tab:nas-timer-evolution} summarizes these milestones.

\begin{table*}[!t]
\centering
\caption{3GPP milestones in NTN-related NAS timer standardization.}
\normalfont\footnotesize
\definecolor{nasRowHead}{RGB}{209,221,236}
\definecolor{nasMatrixBG}{RGB}{245,247,251}
\definecolor{nasMatrixRule}{RGB}{89,133,185}
\setlength{\tabcolsep}{5pt}
\setlength{\extrarowheight}{3pt}
\setlength{\arrayrulewidth}{0.6pt}
\setlength{\aboverulesep}{0pt}
\setlength{\belowrulesep}{0pt}
\renewcommand{\arraystretch}{1.4}
\arrayrulecolor{nasMatrixRule}
\begin{tabularx}{\textwidth}{>{\columncolor{nasRowHead}\raggedright\arraybackslash\bfseries}X
*{6}{>{\columncolor{nasMatrixBG}\centering\arraybackslash}m{1.08cm}}}
% A single left header cell avoids seams across its text.
\textbf{Standardization Milestone} &
\multicolumn{6}{@{}c@{}}{%
\begin{tabular}[c]{*{6}{>{\columncolor{nasMatrixBG}\centering\arraybackslash}m{1.08cm}}}
\multicolumn{2}{>{\columncolor{nasMatrixBG}}c|}{\textbf{Study Phase}}
& \multicolumn{3}{>{\columncolor{nasMatrixBG}}c|}{\textbf{Normative Evolution}}
& \multicolumn{1}{>{\columncolor{nasMatrixBG}}c}{\textbf{Proposal}} \\
\midrule[0.6pt]
\textbf{Rel-15} & \textbf{Rel-16} & \textbf{Rel-17} &
\textbf{Rel-18} & \textbf{Rel-19} & \textbf{Rel-20} \\
\end{tabular}%
} \\
\midrule[0.6pt]
Study on NR support for NTN & $\checkmark$ & & & & &\\
Solutions study for NR NTN & & $\checkmark$ & & & &\\
Normative NR NTN specifications & & & $\checkmark$ & & &\\
Satellite-specific NAS timer values & & & $\checkmark$ & & &\\
Satellite-RAT-type-based timer selection & & & $\checkmark$ & & &\\
5GSM reference correction and timer-handling clarification & & & & $\checkmark$ & &\\
Explicit NR(LEO) no-extension rule & & & & & $\checkmark$ &\\
Inter-orbit timer update & & & & & & $\triangle$ \\
\end{tabularx}
\arrayrulecolor{black}
\par\vspace{4pt}
\begin{minipage}{\textwidth}
\scriptsize
$\checkmark$ denotes a completed study or an incorporated normative change; 
$\triangle$ denotes a proposed change not incorporated into TS~24.501.
\end{minipage}
\label{tab:nas-timer-evolution}
\end{table*}

Rel-18 corrected a 5GSM timer-table reference and clarified timer-value handling for satellite NG-RAN access. Rel-19 made explicit that satellite-specific timer extensions do not apply to NR(LEO) \cite{3gpp_cr6435}. These incorporated changes refined timer references and selection rules without defining separate NTN-specific 5GMM or 5GSM procedures. By contrast, CR 7143 rev. 1 proposed updating running NAS timers during inter-orbit mobility, but it was postponed at CT1\#160 and did not alter the normative baseline \cite{3gpp_cr7143}.

\subsection{NAS Timer Functions, Ownership, and Procedure Boundaries}

TS~24.501 specifies NAS timer behavior for 5GS mobility management (5GMM) and 5GS session management (5GSM). Depending on the procedure, 5GMM timers are maintained by the UE or AMF, whereas 5GSM timers are maintained by the UE or Session Management Function (SMF). In the 5G Core (5GC) architecture shown in Fig.~\ref{fig:ntn-nas-overview}, the AMF terminates 5GMM signaling and relays 5GSM messages between the UE and SMF. The Authentication Server Function (AUSF) and Unified Data Management (UDM) support authentication and subscription-data management, respectively \cite{3gpp_ts23502}.

For a response-supervision timer, the standardized start and stop events delimit the interval observed by its timer owner, namely the UE or network function that maintains it. The elapsed time includes message transport and any intervening processing, while expiry triggers the recovery actions specified for the procedure. Satellite-path delay is therefore relevant to the exchanges occurring within this interval. Backoff, periodic-update, and validity timers govern subsequent attempts, periodic registration, or the lifetime of protocol state. Their duration is tied to policy and protocol state instead of response arrival in an ongoing exchange.

Registration illustrates how different timers cover distinct portions of a procedure. The UE starts T3510 when sending REGISTRATION REQUEST and normally stops it upon receiving REGISTRATION ACCEPT or REGISTRATION REJECT. If authentication or security mode control is performed, the AMF uses T3560 to await the response to AUTHENTICATION REQUEST or SECURITY MODE COMMAND, respectively, with a separate activation for each exchange. When the UE is required to return REGISTRATION COMPLETE, the AMF starts T3550 upon transmitting REGISTRATION ACCEPT and stops it upon receiving REGISTRATION COMPLETE.

\subsection{Satellite-Access Timer Selection in TS~24.501}

TS~24.501 provides satellite-specific extended values for selected 5GMM and 5GSM timers. These extensions apply to NR(MEO) and NR(GEO), whereas NR(LEO) retains the normal 5GS values \cite{3gpp_ts24501}. For the registration timers considered here, T3510 uses 15 s normally and 27 s with the extension, while T3550 and T3560 use 6 s normally and 11 s with the extension.

For 5GMM, the UE and AMF select the applicable values for their respective timers based on the satellite NG-RAN RAT type determined from lower-layer information. The UE also uses this access information for 5GSM timer selection, whereas the SMF applies the specified extended values when their use is indicated by the AMF. Each applicable value is determined at the start of the corresponding NAS procedure and is not recalculated until that procedure is completed, restarted, or aborted \cite{3gpp_ts24501}.

\section{Limitations of NAS Timer Selection by Satellite RAT Type}

NAS timer suitability depends on the delay accumulated between its standardized start and stop events. A timer-centric view of this interval reveals variation within a LEO network that RAT-type-based selection does not distinguish.

\subsection{Timer-Protected Intervals in NAS Procedures}

An individual response-supervision timer is associated with a defined message exchange. The timer owner starts it at the event specified by the procedure and stops it when the expected message is received or another standardized stop condition is met. These NAS events delimit the interval that contributes directly to the elapsed timer value. Earlier activity can still shape the interval through residual queues and protocol state.

Because different timers have different managing entities and start--stop events, they can protect different portions of the same control-plane procedure. A UE-owned request timer can include request delivery, the network-side processing required before a response is produced, and response delivery to the UE. By contrast, an AMF-owned timer that starts with a network command or acceptance message covers the return exchange up to the corresponding UE response. For 5GSM procedures, the protected interval can additionally include transport and processing between the AMF and SMF because UE--SMF messages are relayed through the AMF.

\begin{figure*}[!t]
\centering
\includegraphics[width=0.92\textwidth]{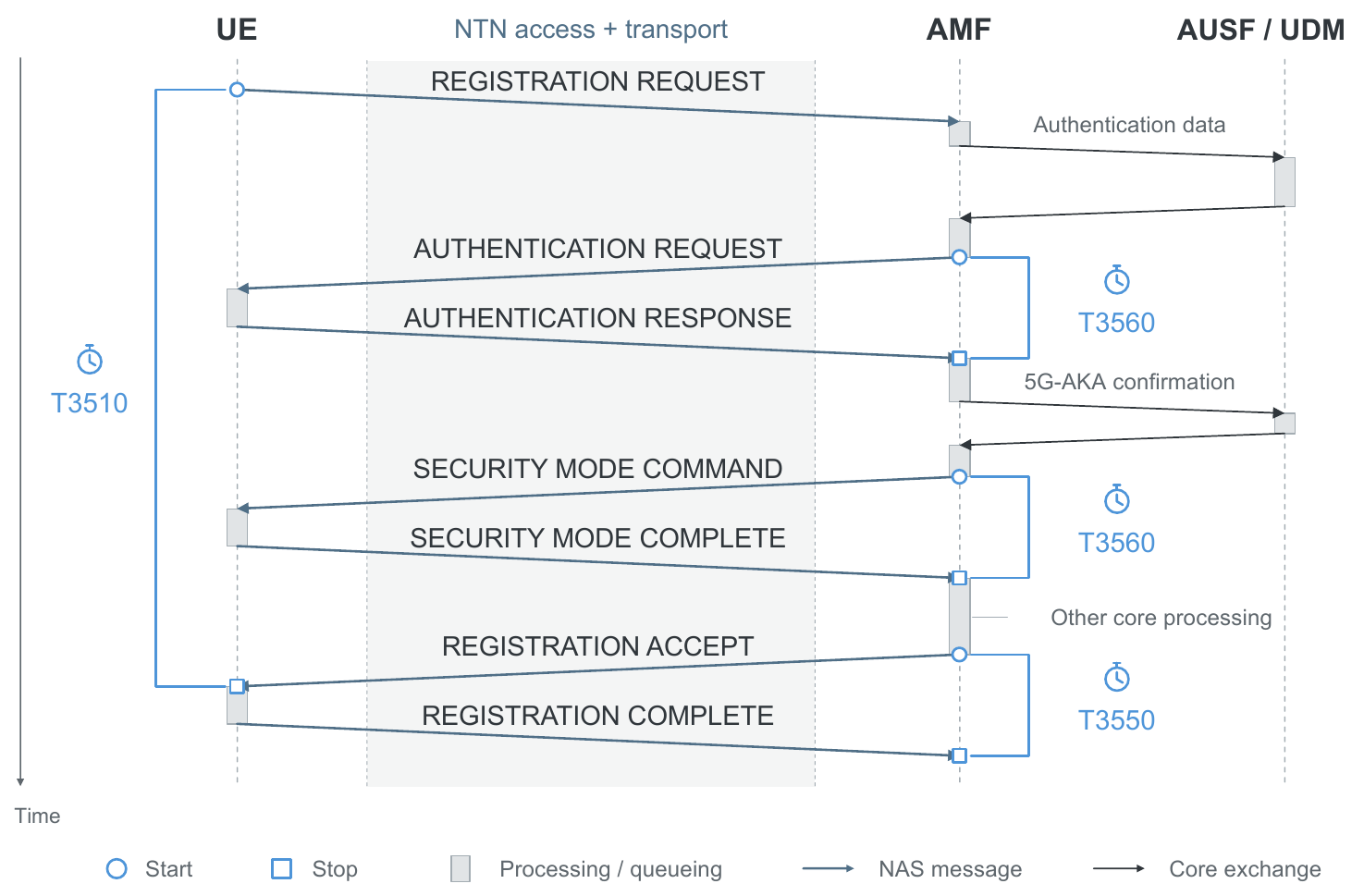}
\caption{NAS timer intervals along a successful 5GS registration path over an NTN.}
\label{fig:timer-protected-interval}
\end{figure*}

Figure~\ref{fig:timer-protected-interval} illustrates how different response-supervision timers cover distinct portions of a successful registration path. T3510 spans from the REGISTRATION REQUEST to the REGISTRATION ACCEPT, including the authentication and security-mode exchanges. The two T3560 intervals are successive AMF-side instances that supervise these exchanges separately. T3550 covers only the exchange from REGISTRATION ACCEPT to REGISTRATION COMPLETE. The vertical spacing is schematic and not proportional to delay.

\subsection{Delay Components Along the Protected Control Path}

Within a timer-protected interval, delay can accumulate in both message directions and during the processing required before the stopping message is produced. For satellite access, propagation can span the service link, one or more ISLs, and the feeder link. Its value depends on UE--satellite--gateway geometry and the selected satellite route. Payload architecture, gateway selection, and network deployment determine which satellite and terrestrial segments are traversed.

Propagation is one part of the interval. Radio scheduling and lower-layer recovery affect message delivery between the UE and NG-RAN. Terrestrial transport can contribute forwarding and queueing delay, while the AMF, SMF, and other 5GC functions can contribute processing and queueing before the expected response is produced.

These factors differ in both variability and observability. UE--satellite geometry changes predictably with location and orbital motion, whereas route changes, lower-layer recovery, and network load can vary on shorter or less predictable time scales. A UE can obtain location and radio-access information but normally has no direct view of core-network queueing. An AMF or SMF can observe local processing conditions but may not know the current satellite route or the lower-layer recovery experienced by the UE. A factor can be used directly as a timer-selection input only when the applicable timer value is determined at procedure start.

\subsection{Implications of Timer Selection by Satellite RAT Type}

The current framework uses the satellite NG-RAN RAT type to select a predefined timer value when a NAS procedure starts. The selected value remains unchanged for that procedure instance. This mapping provides predefined selection rules and preserves the standardized timer state machine.

This timer-centric view makes the granularity of the mapping explicit. NAS timer values remain timer-specific, but the satellite-related selection input is the satellite NG-RAN RAT type. For a selected timer, NR(LEO) instances therefore use the same normal 5GS value even when the satellite route, gateway, terrestrial transport, or processing conditions differ. Two instances of the same timer can consequently have different protected intervals under the same value. Two timers using the same control path can also include different delay components because their owners and stopping messages differ. Timer refinement must therefore align its inputs with the protected exchange.

These observations identify three groups of candidate inputs for timer refinement: geometry and route, access behavior, and transport and processing conditions. Their use is constrained by information availability and timer ownership, which determine where selection can occur and whether assistance is needed.

\section{Directions for Fine-Grained NAS Timer Configuration}

Fine-grained NAS timer configuration distinguishes path and processing conditions that are aggregated under the current satellite NG-RAN RAT-type mapping. Two key aspects can be considered: configuration granularity and information scope.

\subsection{Configuration Granularity}

The current satellite NG-RAN RAT-type mapping provides the baseline for comparison. For a given timer, procedure instances associated with the same RAT type receive the same predefined value. The mapping supports common provisioning and testing, but it does not distinguish the path and processing conditions encountered within the NR(LEO) RAT type.

At an intermediate granularity, a candidate mechanism assigns a common timer value to procedure instances that share observable operating conditions. The grouping can be based on service-link geometry, route length, gateway attachment, load range, or combinations of these inputs. Each group-level value must account for unobserved conditions and response-time variation within the group. Mapping the groups to a finite set of predefined values retains testable configurations while distinguishing conditions not captured by RAT-type selection. AstroTimer~\cite{AstroTimer2026} exemplifies this granularity by deriving one common value for a given load from a delay model evaluated at a reference path.

At the finest granularity considered here, a timer value could be calculated separately for each UE at the start of a procedure instance. Path geometry can be estimated when UE location and satellite ephemeris are available. Gateway attachment and route information extend the estimate across the feeder link and any ISLs, while processing estimates account for delay at the relevant network functions. Location-dependent layer-2 timer design~\cite{Sheemar2024_LocationTimers} and UE-specific NAS delay modeling~\cite{Liu2026_LocationAwareNAS} provide precedents for using these inputs. The selected value remains fixed for the procedure instance, even when the underlying conditions later change.

Group-level and per-instance values both require calibration against the relevant timer-protected interval. Response-arrival distributions characterize premature-expiry risk, whereas retransmission and failure-detection delay after message loss should be evaluated under common loss, retry, and backoff assumptions.

\subsection{Information Scope and Standardization Implications}

Information scope is independent of configuration granularity. Owner-local selection uses context already available at the entity that manages the timer. A UE can use its location and radio-access state, while an AMF or SMF can use locally observed processing conditions. Satellite route, gateway, lower-layer recovery, and remote queueing information may reside at other entities. Cross-entity assistance could make these inputs available to the timer owner before selection.

Both local and assisted selection remain subject to the TS~24.501 timer lifecycle~\cite{3gpp_ts24501}. The information used for selection must be available to the timer owner when the corresponding NAS procedure starts. The selected value remains unchanged until the procedure is completed, restarted, or aborted.

The timing of assistance therefore depends on timer ownership and the corresponding procedure. AMF queueing before REGISTRATION ACCEPT contributes to the interval observed by the UE under T3510, so information used for T3510 selection must reach the UE before it sends REGISTRATION REQUEST. For T3550, the AMF determines the applicable value when the registration procedure starts. The timer is subsequently activated upon transmission of REGISTRATION ACCEPT and protects the return of REGISTRATION COMPLETE. The same reasoning applies to other NAS timers based on their respective procedure boundaries.

Owner-local selection avoids new cross-entity signaling, but a mechanism that changes the applicable TS~24.501 value would still require normative support. Standardized profiles would require common input definitions, range boundaries, value mappings, and conformance test conditions. Cross-entity assistance would additionally require identified information producers and consumers, binding to the relevant access and procedure instance, a validity period, capability handling, and fallback behavior. The applicable TS~24.501 value would provide a natural fallback when the required information is unavailable or invalid.

\section{Case Study of NAS Timer Configuration in LEO Networks}

This case study examines how NAS timer configurations affect registration over heterogeneous LEO paths. It compares a fixed NR(LEO) baseline, a group level configuration, and a per-UE configuration under registration bursts and message loss.

\subsection{Evaluation Scenario and Timer Configurations}

The simulation builds on the registration framework in~\cite{Liu2026_LocationAwareNAS}. It considers a terrestrial AMF reached through a single gateway and a 14-satellite relay chain. A geometry model at an orbital altitude of 550 km provides path estimates for UEs traversing 0–5 or 8 ISLs. In the zero-hop case, the serving satellite connects directly to the gateway through the feeder link. Registration bursts contain 3,000, 4,000, or 5,000 UEs, all initiating registration simultaneously. For each burst size, UE locations and routes are identical across configurations.

The NR(LEO) baseline, labeled 3GPP in the figures, uses T3510=15~s and T3550=T3560=6~s for all UEs. The AstroTimer-based configuration~\cite{AstroTimer2026} selects a common value per timer and load using a reference path. The location-aware (LA) method~\cite{Liu2026_LocationAwareNAS} uses per-UE path estimates, queue estimates based on modeled arrival order, and estimated path reliability. Both methods select T3510, T3550, and T3560 before registration and keep the values fixed across each UE's attempts. All configurations use T3511=10 s between attempts and allow at most five attempts per UE. The AMF may retransmit AUTHENTICATION REQUEST and REGISTRATION ACCEPT up to four times each within an attempt.

egistration is counted as successful when the UE receives REGISTRATION ACCEPT. The T3510 expiry rate measures the fraction of T3510 activations that expire across all registration attempts. Mean registration time covers the interval from the first REGISTRATION REQUEST to this event, including retries and intervening waits, for successful UEs only.

\begin{figure*}[!t]
\centering
\includegraphics[width=0.98\textwidth]{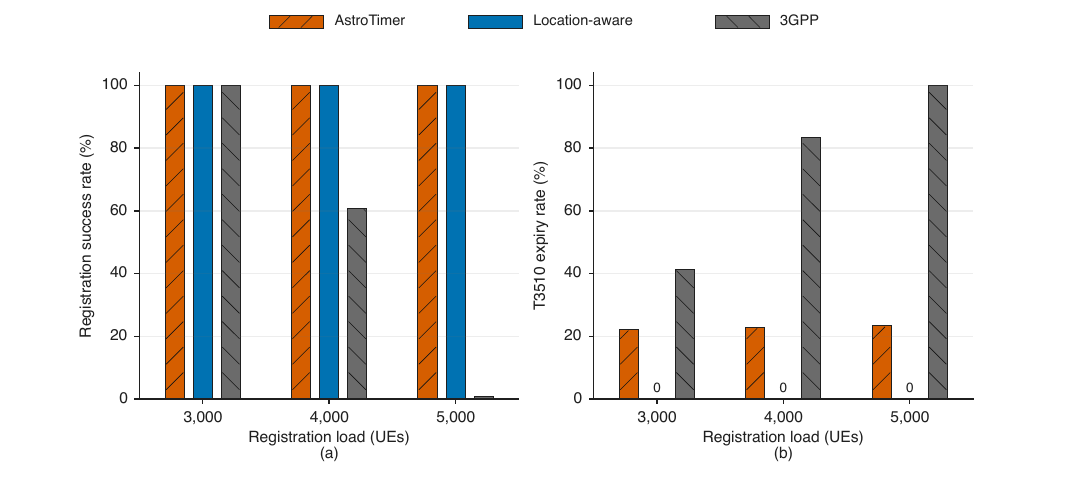}
\caption{Registration success rate and T3510 expiry rate under increasing registration load.}
\label{fig:case-reliability}
\end{figure*}

\begin{figure*}[!t]
\centering
\includegraphics[width=0.98\textwidth]{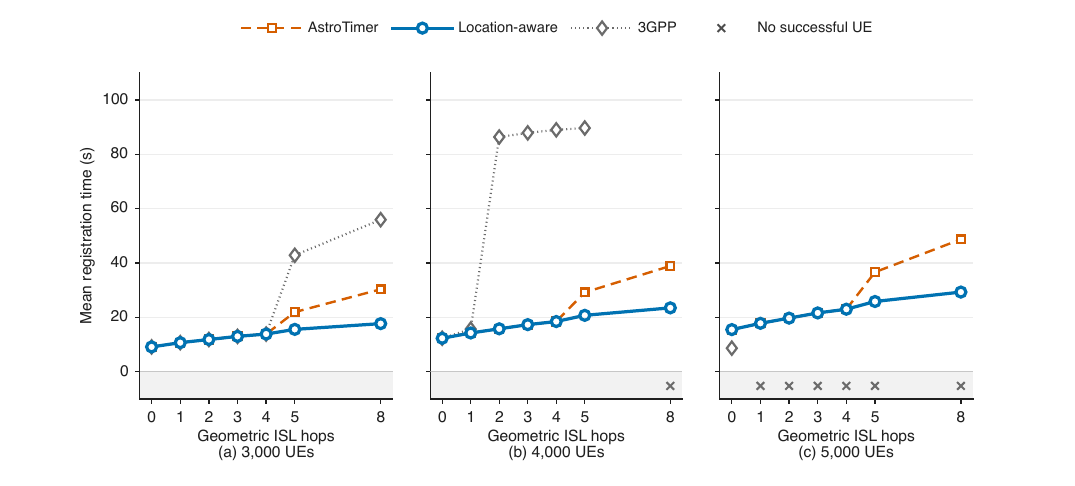}
\caption{Mean registration time across ISL hop counts.}
\label{fig:case-registration-time}
\end{figure*}

\subsection{Premature Expiry Without Message Loss}

Figure~\ref{fig:case-reliability} compares registration success and T3510 expiry as the load increases. The NR(LEO) baseline registers every UE at 3,000 UEs, although its T3510 expiry rate is already 41.27\%. At 4,000 and 5,000 UEs, the expiry rate rises to 83.35\% and 99.83\%, while success falls to 60.78\% and 0.82\%, respectively. No eight-hop UE registers at 4,000 UEs, and successful registrations are confined to zero-hop paths at 5,000 UEs. AstroTimer and LA register every UE at all three loads. AstroTimer still records T3510 expiry rates of 22.12–23.29\%, whereas LA records no expiry and completes every registration in one attempt.

Figure 4 shows identical mean registration times for AstroTimer and LA over zero- to four-hop paths, where neither experiences T3510 expiry. Their results diverge at five and eight hops. Under AstroTimer, T3510 expires on the first attempt for every eight-hop UE and about half of the five-hop UEs. These UEs succeed on their second attempt after the intervening T3511 wait. By avoiding these retries, LA reduces mean registration time relative to AstroTimer by 20.2–21.0\% overall, 28.9–29.5\% on five-hop paths, and 39.6–41.6\% on eight-hop paths.

\begin{figure*}[!t]
\centering
\includegraphics[width=0.98\textwidth]{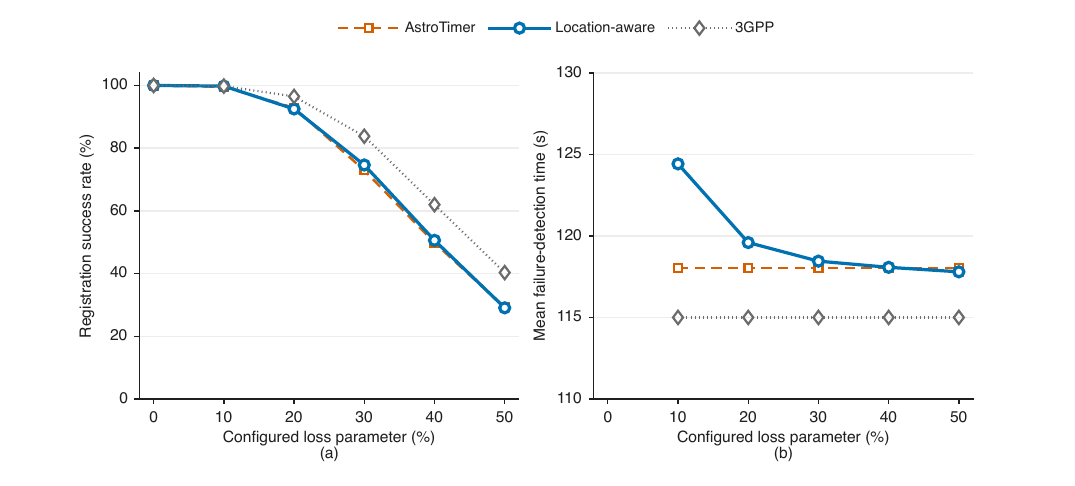}
\caption{Registration success rate and failure-detection time under increasing message loss.}
\label{fig:case-loss}
\end{figure*}

\subsection{Failure Detection Under Message Loss}

The same methods are evaluated at 3,000 UEs, where all three register every UE without message loss. A configured loss parameter ranging from 0 to 50\% is mapped to independent packet-loss probabilities on the service, inter-satellite, and feeder links. Figure~\ref{fig:case-loss}(a) shows declining registration success under all three methods as the parameter increases. At 10\%, their success rates differ by less than 0.3 percentage points. From 20\% onward, the baseline registers more UEs than AstroTimer or LA. At 50\%, the success rates are 40.30\%, 29.30\%, and 29.07\%, respectively.

Figure~\ref{fig:case-loss}(b) reports the mean failure-detection time for the unsuccessful UEs of each configuration. It spans the first REGISTRATION REQUEST to the fifth T3510 expiry, including four intervening T3511 waits. The mean remains at 115 s for the baseline and approximately 118 s for AstroTimer across the nonzero loss settings. LA selects shorter per-UE values as modeled path reliability decreases. Its observed mean failure-detection time falls from 124.4 s at 10\% to 117.8 s at 50\%, close to AstroTimer.

The loss-free results show that a 15 s T3510 can expire before delayed responses arrive. Under message loss, the baseline's shorter T3550 and T3560 values trigger earlier retransmissions when expected responses are absent. This helps explain its higher success rates at the larger loss settings. Its shorter T3510 also limits the time spent on unsuccessful attempts.

Within this case study, per-UE selection removes the long-path expiries observed with the tested common configuration, while longer waits can delay recovery after loss. The benefit of finer-grained selection therefore depends on calibration for both delayed responses and message loss. Registration success and failure-detection time should be assessed together when choosing timer values.

\section{Conclusion and Outlook}

NAS timer configuration over satellite access must account for heterogeneous control-path delays while preserving standardized procedure behavior. The current satellite NG-RAN RAT-type mapping does not distinguish route and processing variations within NR(LEO). Finer-grained selection should therefore reflect each timer’s protected exchange and the information available to its owner.

The case study illustrates the trade-off between premature expiry and recovery after message loss. Per-UE selection avoids the long-path expiries observed with the common configuration, while shorter timer values enable earlier retransmission after loss. Timer calibration should therefore consider registration success and registration time alongside failure-detection time.

Future specification work should address configuration granularity and information scope as separate design dimensions. Deployment evidence should guide the choice between shared profiles and per-instance values. For both owner-local and assisted selection, specifications need common input definitions, selection rules, and conformance tests. Assisted selection additionally requires rules for information availability at procedure start, validity, and fallback behavior.

\section{References}
\bibliographystyle{IEEEtran}
\bibliography{references}

\section{Acknowledgment}
We acknowledge the support of the Natural Sciences and Engineering Research Council of Canada (NSERC), [funding reference number RGPIN-2022-03364].

%\section{Biographies}
% \vspace*{-30pt}
\begin{IEEEbiographynophoto}
{Cheng Liu} (liuc13@myumanitoba.ca) received the B.Eng. and M.Eng. degrees from Hubei University of Technology and Wuhan University of Technology, respectively. He is currently pursuing the Ph.D. degree with the Department of Electrical and Computer Engineering, University of Manitoba, Winnipeg, MB, Canada. His research interests include 6G wireless communications, non-terrestrial networks, and the Internet of Things.
\end{IEEEbiographynophoto}

\begin{IEEEbiographynophoto} %150 words
{Peng Hu} [Senior Member] (peng.hu@umanitoba.ca) received the Ph.D. degree in Electrical Engineering from Queen’s University, Canada. He is currently an Associate Professor in the Department of Electrical and Computer Engineering at the University of Manitoba, an Adjunct Associate Professor in the Cheriton School of Computer Science at the University of Waterloo, Canada, and a member of the Waterloo Data and AI Institute. He serves as Secretary of the IEEE ComSoc's Satellite and Space Communications Technical Working Group, Associate Editor for the \textit{IEEE Transactions on Aerospace and Electronic Systems} and \textit{IEEE Open Journal of the Communications Society}, and West Area Chair of Region 7. He previously served as Co-Chair of the Technology Working Group of the IEEE LEO Satellites \& Systems Project and as a member of the IEEE Sensors Council Standards Committee. His research interests include space-air-ground integrated networks, non-terrestrial networks, space sustainability, and autonomous networking.

%received the Ph.D. degree in Electrical Engineering from Queen's University, Canada. He is currently an Associate Professor in the Department of Electrical and Computer Engineering at the University of Manitoba, an Adjunct Associate Professor in the Cheriton School of Computer Science at the University of Waterloo, and a member of the Waterloo Data and AI Institute. He serves as the Secretary of the Satellite and Space Communications Technical Working Group of the IEEE Communications Society, an Associate Editor of the \textit{IEEE Transactions on Aerospace and Electronic Systems} and \textit{IEEE Open Journal of the Communications Society}. He also serves at the West Area Chair of IEEE Region 7, and  has previously served as the Co-Chair of the Technology Working Group of the IEEE LEO Satellites \& Systems Project and Associate Editor of the \textit{Canadian Journal of Electrical and Computer Engineering}, a voting member of the IEEE Sensors Council Standards Committee, and a member of organizing and technical program committees for various international conferences and industry consortia, including IEEE ICC, IEEE GLOBECOM, IEEE PIMRC, IEEE WiSEE, and IEEE AINA. His current research interests include space-air-ground integrated networks, non-terrestrial networks, space sustainability, autonomous networking, and the Internet of Things.
\end{IEEEbiographynophoto}

\end{document}